\documentstyle[fixa4,fleqn,epsfig,12pt,rotating]{article}
\def\3{\ss}                                                                                        
                                                                           
\renewcommand{\author}{ }                                                                          
\begin{document}
\vspace{1 cm}
%
%---------- Abbreviations -------------
%
\newcommand{\bec}       {\begin{center}}
\newcommand{\eec}       {\end{center}}
\newcommand{\qsq}       {\mbox{$Q^{2}$}}
\newcommand{\jpsi}      {\mbox{$J/\psi$}}
\newcommand{\Acce}	{\mbox{$\mathcal{A}$}}
\newcommand{\Lumi}	{\mbox{$\mathcal{L}$}}
\newcommand{\BR}	{\mbox{$\mathcal{B}$}}
\newcommand{\mv}        {\mbox{$m_{V}$}}
\newcommand{\wgp}       {\mbox{$W_{\gamma p}$}}
\newcommand{\zv}        {\mbox{$z_{\rm vert}$}}
\newcommand{\sleq}      {\raisebox{-.6ex}{${\textstyle\stackrel{<}{\sim}}$}}
\newcommand{\sgeq}      {\raisebox{-.6ex}{${\textstyle\stackrel{>}{\sim}}$}}
\newcommand{\Gev}       {\mbox{${\rm GeV}$}}
\newcommand{\Gevsq}     {\mbox{${\rm GeV}^2$}}
\newcommand{\x}         {\mbox{${\it x}$}}
\newcommand{\smallqsd}  {\mbox{${q^2}$}}
\newcommand{\ra}        {\mbox{$ \rightarrow $}}
%
% ---- commands from paul -----
%
\def\ctr#1{{\it #1}\\\vspace{10pt}}
\def\si{{\rm si}}
\def\Si{{\rm Si}}
\def\Ci{{\rm Ci}}
\def\px{p_{_{x}}}
\def\py{p_{_{y}}}
\def\pz{p_{_{z}}}
\def\yjb{y_{_{JB}}}
\def\xjb{x_{_{JB}}}
\def\qjb{\qsq_{_{JB}}}
\def\gap{\hspace{0.5cm}}
\renewcommand{\thefootnote}{\arabic{footnote}}
%
% --------  Title, Date and Authors
%
\begin{flushright}{\large DESY-97-060} \end{flushright}
\vspace{1cm}
\bec{\Large\bf
    Measurement of Elastic $J/\psi$ Photoproduction at HERA
     }\eec
\vspace{1cm}
\vspace{1.0cm}
\bec{\bf 
    ZEUS Collaboration
    }\eec
\vspace{3cm}
%
% -------- Abstract -------------------
%
\begin{abstract}
The reaction $\gamma~p \rightarrow \jpsi~p$ has been studied in $ep$
interactions using the ZEUS detector at HERA.
The cross section for elastic \jpsi~photoproduction has been measured as a 
function of the photon-proton centre of mass energy $W$ in the range 
$40 < W < 140$~GeV at a median photon virtuality 
\qsq~of~$5 \times 10^{-5}$~GeV$^2$.
The photoproduction cross section,
$\sigma_{\gamma p \rightarrow J/\psi p}$, is observed to rise steeply with
$W$. 
A fit to the data presented in this paper to determine the parameter
$\delta$ in the form
$\sigma_{\gamma p \rightarrow J/\psi p} \propto W^{\delta}$ yields the
value $\delta = 0.92 \pm 0.14 \pm 0.10$. 
The differential cross section ${d\sigma}/{d|t|}$ is presented over the range
$|t| < 1.0$~GeV$^2$ where $t$ is the square of the four-momentum exchanged at
the proton vertex.  
${d\sigma}/{d|t|}$ falls exponentially with a slope parameter of
$4.6\pm0.4^{+0.4}_{-0.6}$~GeV$^{-2}$.
The measured decay angular distributions are consistent with
$s$-channel helicity conservation.
\end{abstract}
%
%--------  Reset page counter and goto next page
%
\setcounter{page}{0}
\thispagestyle{empty}
\pagenumbering{Roman}                                                                              
\newpage
%===================================================================
%                                                                   
%  MEMBER NAME  AUTH44 (ZEUS)     M  TEX                            
%                                                                                                  
%  JH.: transformed to a format, which is suited as input for                                      
%       CONVERT, which automatically creates author-indices                                        
%                                                                                                  
%  Don't remove lines starting with a percent sign %,                                              
%  CONVERT may need them urgently !                                                                
%                                                                                                  
%=====================================================================                             
%
\begin{center}                                                                                     
{                      \Large  The ZEUS Collaboration              }                               
\end{center}                                                                                       
  J.~Breitweg,                                                                                     
  M.~Derrick,                                                                                      
  D.~Krakauer,                                                                                     
  S.~Magill,                                                                                       
  D.~Mikunas,                                                                                      
  B.~Musgrave,                                                                                     
  J.~Repond,                                                                                       
  R.~Stanek,                                                                                       
  R.L.~Talaga,                                                                                     
  R.~Yoshida,                                                                                      
  H.~Zhang  \\                                                                                     
 {\it Argonne National Laboratory, Argonne, IL, USA}~$^{p}$
% ...
\newpage                                                                                           
% ...
$^{\    2}$ now at Univ. of Salerno and INFN Napoli, Italy \\                                      
% ...
$^{  23}$ now at ATLAS Collaboration, Univ. of Munich \\                                           
% ...
                                                           %                                       
                                                           %                                       
% \par         % if index listing & table fit to 1 page, put gap here                              
\newpage   % alternatively: go to newpage, if page is too small                                    
                                                           %                                       
% \institute_references_start    % do not touch or move this line !                                
                                                           %                                       
\begin{tabular}[h]{rp{14cm}}                                                                       
$^{a}$ &  supported by the Natural Sciences and Engineering Research                               
          Council of Canada (NSERC)  \\                                                            
$^{b}$ &  supported by the FCAR of Qu\'ebec, Canada  \\                                            
$^{c}$ &  supported by the German Federal Ministry for Education and                               
          Science, Research and Technology (BMBF), under contract                                  
          numbers 057BN19P, 057FR19P, 057HH19P, 057HH29P, 057SI75I \\                              
$^{d}$ &  supported by the MINERVA Gesellschaft f\"ur Forschung GmbH,                              
          the German Israeli Foundation, and the U.S.-Israel Binational                            
          Science Foundation \\                                                                    
$^{e}$ &  supported by the German Israeli Foundation, and                                          
          by the Israel Science Foundation                                                         
  \\                                                                                               
$^{f}$ &  supported by the Italian National Institute for Nuclear Physics                          
          (INFN) \\                                                                                
$^{g}$ &  supported by the Japanese Ministry of Education, Science and                             
          Culture (the Monbusho) and its grants for Scientific Research \\                         
$^{h}$ &  supported by the Korean Ministry of Education and Korea Science                          
          and Engineering Foundation  \\                                                           
$^{i}$ &  supported by the Netherlands Foundation for Research on                                  
          Matter (FOM) \\                                                                          
$^{j}$ &  supported by the Polish State Committee for Scientific                                   
          Research, grant No.~115/E-343/SPUB/P03/120/96  \\                                        
$^{k}$ &  supported by the Polish State Committee for Scientific                                   
          Research (grant No. 2 P03B 083 08) and Foundation for                                    
          Polish-German Collaboration  \\                                                          
$^{l}$ &  partially supported by the German Federal Ministry for                                   
          Education and Science, Research and Technology (BMBF)  \\                                
$^{m}$ &  supported by the German Federal Ministry for Education and                               
          Science, Research and Technology (BMBF), and the Fund of                                 
          Fundamental Research of Russian Ministry of Science and                                  
          Education and by INTAS-Grant No. 93-63 \\                                                
$^{n}$ &  supported by the Spanish Ministry of Education                                           
          and Science through funds provided by CICYT \\                                           
$^{o}$ &  supported by the Particle Physics and                                                    
          Astronomy Research Council \\                                                            
$^{p}$ &  supported by the US Department of Energy \\                                              
$^{q}$ &  supported by the US National Science Foundation \\                                       
\end{tabular}                                                                                      
                                                           %                                       
% \institute_references_end     % do not touch or move this line !                                 
%
%--------  Reset page counter and goto next page
%
%\thispagestyle{empty}
\newpage
\setcounter{page}{1}
\pagenumbering{arabic}                                                                              
%
%--------  Introduction and Motivation
%
\section{\bf Introduction}
\label{Sect:Intro}

This paper reports new data on the photoproduction of the \jpsi~
meson using the ZEUS detector at HERA.
It is part of our continuing study of vector meson
($\rho,\omega,\phi,\jpsi$) production in both the photoproduction
\cite{Ref:ZEUSRho, Ref:ZEUSOme, Ref:ZEUSPhi, Ref:ZeusJPsi93}
and the deep inelastic scattering regimes
\cite{Ref:ZEUSDISRho, Ref:ZEUSDISPhi}.
Previous results have established a weak dependence on the
photon-proton centre of mass energy, $W$, of the vector meson
photoproduction cross sections 
($\sigma \propto W^{\delta}$ with $\delta \sim 0.2-0.3$) if there is
no hard scale in the process, as expected from soft diffraction. 
By contrast, the cross sections for elastic $\rho$ and $\phi$
production in deep inelastic scattering
at $5 \sleq Q^2 \sleq 20$~GeV$^2$ exhibit a stronger $W$
dependence ($\sigma \propto W^{\delta}$ with $\delta \sim 0.3-0.6$)
where $Q^2$ sets the hard scale.
In the photoproduction of the \jpsi~meson the mass of the \jpsi~itself
provides the hard scale and the cross section exhibits a strong $W$
dependence ($\sigma \propto W^{\delta}$ with $\delta \sim 1$).
The total virtual photon-proton cross section \cite{Ref:ZEUSF2,
Ref:H1F2} also exhibits a change in energy dependence as $Q^2$
increases beyond $\approx 1$~GeV$^2$. 
Overall, the data illuminate the transition from the soft,
non-perturbative regime to the kinematic region where perturbative 
descriptions become applicable. 

\jpsi~photoproduction has been measured as a function of 
$W$ from threshold to $W \approx 20$~GeV in  
fixed target experiments \cite{Ref:FixedTrgtJPsi, Ref:E401, Ref:E516}
and extended to $W \approx 140$~GeV at HERA \cite{Ref:ZeusJPsi93,
Ref:H1JPsi93, Ref:H1JPsi94}.
A review of the low energy experimental results can be found in
reference \cite{Ref:Review}.
In this paper we extend our earlier study of elastic $J/\psi$
photoproduction \cite{Ref:ZeusJPsi93} to include the
determination of the differential cross section $d\sigma/d|t|$ and the
angular distributions of the decay leptons.
In addition, the six-fold increase in the size of the data sample allows us to
determine the parameter $\delta$ from the data presented here alone.

The \jpsi~was detected via its leptonic (electron pair and muon pair) decay 
modes in the kinematic range $40 < W < 140$~GeV.
After a brief description of the ZEUS detector, the data taking
conditions, the kinematics of elastic $J/\psi$ production at HERA, 
and the event selection are described.
The $W$ dependence of the 
cross section $\sigma_{\gamma p \rightarrow J/\psi p}$,
the $t$ distribution and the decay angular distributions are then
presented.

\section{Experimental Conditions}
\label{Sect:ExptCond}

\subsection{\bf HERA}
\label{Sect:HERA}

During 1994 HERA operated with a proton beam energy of 820~GeV and a
positron beam energy of 27.5~GeV.
In the positron and proton beams 153 colliding bunches were stored
together with 17 unpaired proton bunches and 15 unpaired positron
bunches.
The time between bunch crossings was 96~ns.
The typical instantaneous luminosity was 
$1.5 \times 10^{30}$~cm$^{-2}$~s$^{-1}$.

\subsection{\bf The ZEUS Detector}
\label{Sect:ZEUS}

The main ZEUS detector components used in this analysis are outlined below.
A detailed description of the ZEUS detector can be found elsewhere~
\cite{Ref:BlueBook}.
In the following the ZEUS coordinate system will be used, the $Z$ axis of 
which is coincident with the nominal proton beam axis, the $X$ axis is 
horizontal and points towards the centre of HERA and the $Y$ axis
completes a right handed coordinate system. 
The origin of the coordinate system lies at the nominal interaction point.

The momentum and trajectory of a charged particle were reconstructed using the
Vertex Detector  (VXD) \cite{Ref:VXD} and the Central Tracking
Detector (CTD)~\cite{Ref:CTD}. 
The VXD and the CTD are cylindrical drift chambers which are placed in
the solenoidal magnetic field of 1.43 T produced by a thin
superconducting solenoid. 
The CTD surrounds the VXD and covers the angular
region $15^o < \theta < 164^o$ (where $\theta$ is the polar angle with respect
to the proton direction).

The high resolution uranium-scintillator calorimeter
CAL~\cite{Ref:CAL} surrounding the coil is 
divided into three parts, the forward calorimeter (FCAL), the barrel 
calorimeter (BCAL) and the rear calorimeter (RCAL), which cover polar angles 
from  $2.6^o$ to $36.7^o$, $36.7^o$ to $129.1^o$, and $129.1^o$ to $176.2^o$, 
respectively. 
Each part consists of towers which are longitudinally subdivided into 
electromagnetic (EMC) and hadronic (HAC) readout cells.

The proton remnant tagger (PRT), a set of scintillation counters
surrounding the beam pipe at small forward angles, serves to tag events
with proton dissociation. 
It is situated at $Z=500$~cm and covers the angular range from 6 to 26~mrad.

The muon detectors \cite{Ref:MUON}, situated outside the calorimeter, 
consist of limited streamer tubes (LST) placed both inside and outside
the magnetised iron yoke. 
The inner chambers (BMUI and RMUI) were used to tag the muons from the
\jpsi. 
The BMUI and the RMUI cover the polar angles between
$34^o < \theta < 135^o$ and $134^o< \theta < 171^o$, respectively.

Proton-gas events occuring upstream of the nominal interaction point are
out of time with respect to the $e^{+} p$ interactions and were rejected
by timing measurements made by the scintillation counter arrays Veto Wall,
C5 and SRTD situated along the beam line at $Z=-730$~cm, $Z=-315$~cm,
and $Z=-150$~cm respectively.

The luminosity was determined from the rate of the 
Bethe-Heitler process $e^{+} p \rightarrow e^{+} \gamma p$ where the photon 
was measured by the LUMI calorimeter located in the HERA tunnel
at $Z=-107$~m \cite{Ref:LumiCalc}. 
The luminosity was determined with a precision of 1.5\% for the measurements 
presented below.

\section{\bf Kinematics}
\label{Sect:Kinem}

\begin{figure}[p]
\begin{center}
\epsfig{file=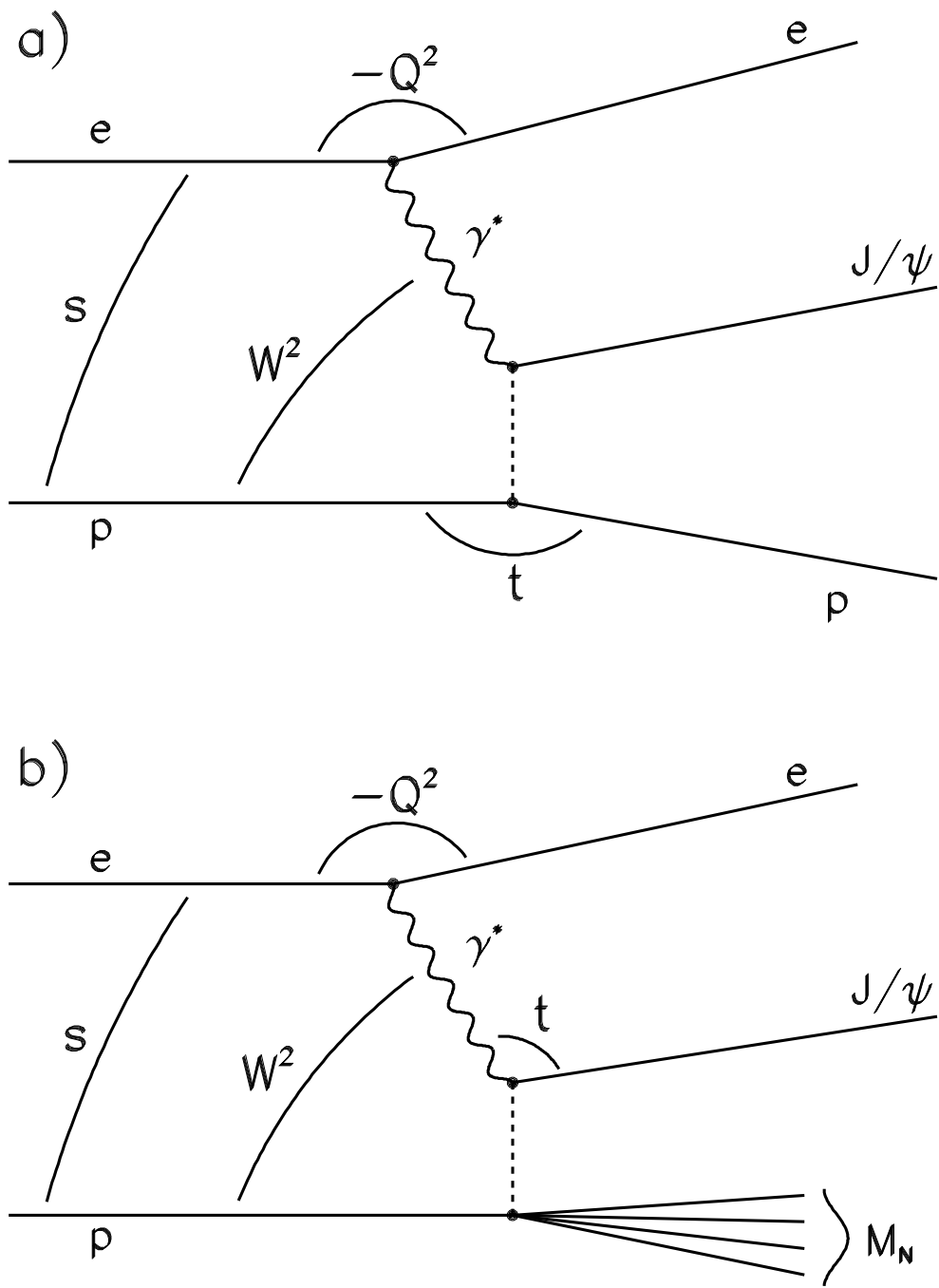,%
      width=13cm%
        }
\caption{
         Schematic diagrams for diffractive \jpsi~electroproduction.
         (a) The mechanism for elastic vector meson production. 
         (b) Proton dissociative \jpsi~photoproduction where the proton 
         dissociates into a hadronic system of invariant mass $M_N$.
         }
\label{Fig:Feynman}
\end{center}
\end{figure}

Figure \ref{Fig:Feynman}a shows a schematic diagram for the reaction:
\begin{equation}
    e^{+}(k)p(P) \rightarrow e^{+}(k')J/\psi(V)p(P'),
\end{equation}
where each symbol in parentheses denotes the four-momentum of the 
corresponding particle.

The kinematics of the inclusive scattering of 
unpolarised positrons and protons are
described by the positron-proton centre of mass energy squared ($s$)
and any two of the following variables
\begin{itemize}
    \item $Q^2=-q^2=-(k-k')^2$, the negative four-momentum squared of the 
         exchanged photon; 
    \item $y=(q\cdot P)/(k\cdot P)$, the fraction of the positron energy 
         transferred to the hadronic final state in the rest frame of the 
         initial state proton;
    \item $W^2 = (q+P)^2= -Q^2+2y(k\cdot P)+M^2_p \approx ys$, the
         centre of mass energy squared of the photon-proton system,
         where $M_p$ is the proton mass.
\end{itemize}

For a complete description of the exclusive reaction 
$e^{+}p \rightarrow e^{+}J/\psi p$
($J/\psi \rightarrow \ell^{+}\ell^{-}$, where $\ell^{+}\ell^{-}$ denotes a
pair of electrons or muons) the following additional variables are
required 
\begin{itemize}
    \item $t = (P-P')^2$, the four-momentum transfer squared at the proton 
         vertex;
    \item the angle between the $J/\psi$ production plane and the positron 
         scattering plane in the photon-proton frame, $\Phi$;
    \item the polar and azimuthal angles, $\theta_h$ and $\phi_h$, of
         the decay leptons in the $J/\psi$ rest frame.
\end{itemize}
In the present analysis, $\Phi$ is not measured because events were
selected in which the scattered positron was not detected.
In such untagged photoproduction events the $Q^2$ value ranges from the 
kinematic minimum $Q^2_{min} = M^2_e y^2/(1-y) \approx 10^{-10}~\rm{GeV^2}$, 
where $M_e$ is the electron mass, to the value at which the scattered positron
starts to be observed in the uranium calorimeter 
$Q^2_{max} \approx 4 \ \rm{GeV^2}$, with a median $Q^2$ of
approximately $5 \times 10^{-5} {\rm~GeV^2}$. 
Since the typical $Q^2$ is small, the photon-proton centre of mass energy can
be expressed as
\begin{equation}
    W^2 \approx 2 (E_{J/\psi} - p_{Z J/\psi}) E_p = 4 E_p E_e y,
\label{Eq:W2Def}
\end{equation}
where $E_p$ and $E_{J/\psi}$ are the laboratory energies of the
incoming proton and the $J/\psi$~and $p_{Z J/\psi}$ is the
longitudinal momentum of the $J/\psi$.
The four-momentum transfer squared, {\it t}, at the proton vertex 
for $Q^2 = Q^2_{min}$ is given by
\begin{equation}
    {\it t} = (q-V)^2 \approx -p^2_{T J/\psi},
\end{equation}
where $p_{T J/\psi}$ is the momentum of the \jpsi~transverse to the beam 
axis.
Non-zero values of $Q^2$ cause $t$ to differ from $-p^2_{T J/\psi}$ by less 
than $Q^2$.
A correction is applied to the $p^2_{T J/\psi}$ distribution 
to correct for this effect as described in section 
\ref{Sect:t_Dist}~\cite{Ref:ZEUSRho}.

\section{\bf Trigger}
\label{Sect:Trigger}

ZEUS uses a three-stage trigger system \cite{Ref:BlueBook}.
The electron and muon pair triggers are outlined below, followed by a summary
of trigger requirements common to both channels.

\vskip 0.5cm
{\parindent 0pt
{\bf Electron Channel}
\vskip 0.5cm

The First Level Trigger (FLT) required 1, 2 or 3 track segments to be
found in the CTD, with at least one segment pointing to the interaction region.
The sum of all the energy deposited in the EMC section of the calorimeter
was required to exceed 0.66~GeV.
In addition, either the total energy in the calorimeter
had to be greater than 2~GeV or the total energy in FCAL (ignoring the cells 
closest to the beam pipe) had to be greater than 2.5~GeV.
}

The Second Level Trigger (SLT) required the total energy in the HAC 
section of the calorimeter to be less than 1~GeV and the total energy in the 
EMC section to be greater than 1.5~GeV.
The ratio of HAC to EMC energy in RCAL and BCAL 
separately had to be less than 0.1 or the HAC energy had to be less than 
0.2~GeV.

The Third Level Trigger (TLT) matched tracks measured in the CTD to 
electromagnetic energy deposits in the calorimeter.
A cluster of contiguous cells, each with an energy of at least 0.3~GeV, was 
defined as electromagnetic if more than 90\% of the total 
cluster energy was contained in EMC cells.
An electron candidate was defined as a track with momentum transverse to the 
beam direction in excess of 0.4~GeV passing within 30~cm of the centre of an 
electromagnetic cluster.
At least two electron candidates of opposite charge were required.
At the distance of closest approach the separation between the two
tracks was required to be less than 7~cm.
An event was kept if the invariant mass of any pair exceeded 2~GeV.

\vskip 0.5cm
{\parindent 0pt
{\bf Muon Channel}
\vskip 0.5cm

At the FLT, track segments had to be found in the inner barrel muon
chambers (BMUI) accompanied by a reconstructed energy deposition of at
least 0.464 GeV in a CAL trigger tower.
Note that on average a muon produces a visible signal of 0.8~GeV in a
trigger tower. Alternatively, hits had to be found in the RMUI
chambers accompanied by a reconstructed energy deposit of at least
0.464 GeV in an RCAL trigger tower \cite{Ref:BlueBook}.
At least one and no more than five track segments had to be found in the 
CTD, with at least one pointing to the interaction region.
}

No requirements were imposed at the SLT.

At the TLT a muon candidate was formed when a track found 
in the CTD matched a cluster of energy in the calorimeter consistent with 
the passage of a minimum ionising particle (m.i.p.) and a
track in the inner muon chambers.
An event containing a muon candidate for which $\theta>147^o$ was accepted if 
the momentum exceeded 1~GeV.
The transverse momentum of a muon candidate for which $20^o<\theta<147^o$
was required to exceed 1~GeV.

\vskip 0.5cm
{\parindent 0pt
{\bf Common Requirements}}
\vskip 0.5cm

An event was rejected at the FLT if the time of arrival of any signal
observed in the Veto Wall, the C5 counter or the SRTD was inconsistent
with the time of the bunch crossing.
In order to increase the purity of the sample the sum of energy in the
inner ring of FCAL was required to be less than 1.25~GeV.

At the SLT, the total energy in the calorimeter ($E_{Tot} = \Sigma_i E_i$) 
and the $Z$ component of the momentum ($\Sigma p_Z = \Sigma_i E_i
\cos \theta_i$) was calculated.
The sums run over all calorimeter cells $i$ for which the energy, $E_i$, 
deposited in the cell is above threshold and the polar angle at which
the cell is found is denoted by $\theta_i$.  
Beam-gas events were rejected by exploiting the excellent time
resolution of the calorimeter.
In order to remove inclusive beam-gas background in time with the
bunch crossing, an event was rejected if the ratio
${\Sigma p_Z}/{E_{Tot}}$ was greater than 0.96. 

Finally, at the TLT, $E_{Tot}$ and $\Sigma p_Z$ were calculated
again using the CAL energies reconstructed at the TLT, and an
event was accepted if $E_{Tot}-\Sigma p_Z \leq 100$~GeV and 
$\Sigma p_Z/E_{Tot} \leq 0.94$.

\section{\bf Offline Event Selection}
\label{Sect:Event}

To be accepted an event was required to have exactly two tracks of
opposite charge with pseudorapidity, $\eta$, in the range $|\eta| < 1.7$.
Denoting the polar angle of a track by $\theta$, $\eta$ is defined
such that $\eta=-\ln\left(\tan({\theta}/{2})\right)$.
The two tracks were required to fit to a common vertex consistent with 
an $ep$ interaction.
The tracks had to match to clusters of energy in the calorimeter and 
events were rejected if more than 1~GeV was deposited in calorimeter cells not
associated with either of the two tracks.
As shown in equation (\ref{Eq:W2Def}), $W^2$ was determined from the measured
$E_{J/\psi} - p_{Z J/\psi}$ of the decay leptons.
The requirement that the value of $W$ lie in the range $40<W<140$~GeV 
restricted the sample to a region of high acceptance.
Selection criteria specific to the electron and muon channel are described 
below.
\vskip 0.5cm
{\parindent 0pt

{\bf Electron Channel}}
\vskip 0.5cm

The electron sample comes from an integrated luminosity of 
$2.70 \pm 0.04$~pb$^{-1}$. 
The algorithm used to define the electron pair sample at the TLT was reapplied
offline with the final detector calibrations.
The transverse momentum threshold of each of the two oppositely charged
tracks was increased to 0.8~GeV.
In order to reduce contamination from misidentified pions, the energy
of at least one of the electromagnetic clusters matched to the tracks
by the TLT algorithm applied offline was required to be larger than 1~GeV.

\begin{figure}[p]
\begin{center}
\epsfig{file=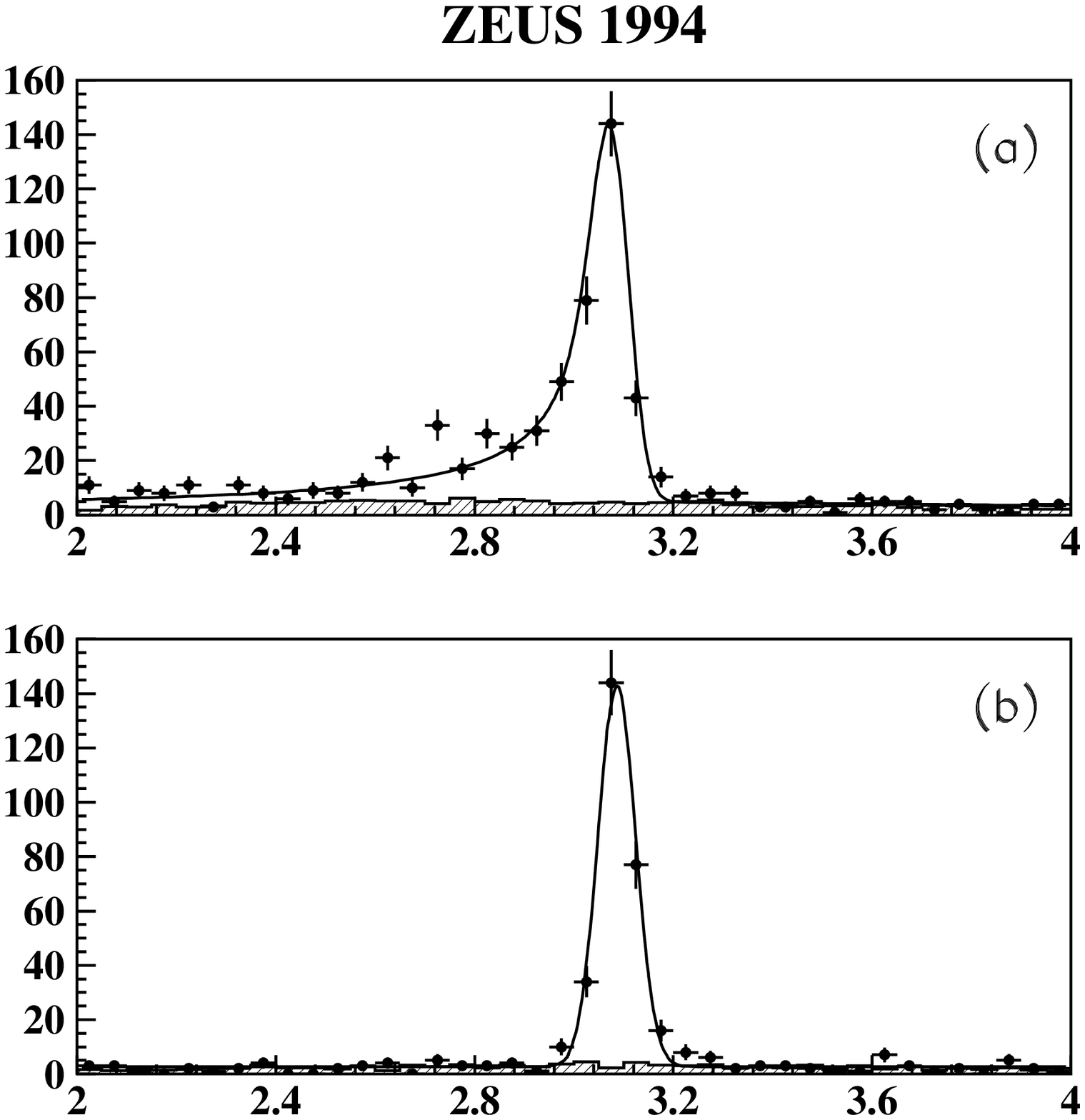,%
      width=15cm}
\put(-140.,204.){\Large{$\bf M_{e^+e^-}$} \large{\bf (GeV)}}
\put(-418.,260.){\begin{rotate}{-90} \large{\bf Events / 50 MeV} \end{rotate}}
\put(-140., 10.){\Large{$\bf M_{\mu^+\mu^-}$} \large{\bf (GeV)}}
\put(-418., 70.){\begin{rotate}{-90} \large{\bf Events / 50 MeV} \end{rotate}}
    \caption{(a) The mass distribution of the events in the electron
            pair sample.
            A clear peak at the \jpsi~mass is observed.
            The solid line shows the result of a fit in which a Gaussian 
            resolution function has been convoluted with a radiative \jpsi~
            mass spectrum and added to a polynomial background.
            (b) The mass distribution for events in the muon pair sample.
            The solid line shows the result of a fit in which a Gaussian 
            resolution function has been added to a flat background 
            function.
            For both the electron and muon channels the contribution of 
            events from the Bethe-Heitler process is shown as the hatched 
            area.
            }
\label{Fig:Mass}
\end{center}
\end{figure}

Figure \ref{Fig:Mass}a shows the mass distribution of the electron pair 
sample.
A clear peak at the \jpsi~mass is observed.
The signal region, $2.85 < M_{e^+e^-} < 3.25$~GeV, contains 392 events.
The cross sections and angular distributions presented below are obtained
by calculating acceptances and background contributions for this range.
The solid line shows an unbinned likelihood fit in which a Gaussian
resolution function has been convoluted with a radiative \jpsi~mass
spectrum and a polynomial describing the background.
The mass estimated by the fit is $3.094 \pm 0.003$~GeV,
the rms width is $33 \pm 4$~MeV, and the number of events
attributable to $J/\psi$ production
estimated by the fit over the mass range $2 < M_{e^+e^-} < 4$~GeV is
$460 \pm 25$. 
\vskip 0.5cm
{\parindent 0pt
{\bf Muon Channel}}
\vskip 0.5cm

The muon sample comes from an integrated luminosity of
$1.87 \pm 0.03$~pb$^{-1}$.
The momentum of each track was required to exceed 1~GeV.
At least one of the two tracks had to match a m.i.p.
cluster in the calorimeter and a track segment in the barrel or rear muon 
chambers.
To remove cosmic ray contamination the calorimeter signals were required to 
be in time with the beam crossing and the distance between the
two tracks must be less than 2~cm at their distance of
closest approach to the beamline.
To further reduce the cosmic ray background the tracks were required
not to be collinear.
This was achieved by calculating the cosine of the angle, $\Omega$,
between the two tracks at the interaction point.
An event was rejected if $\cos \Omega < -0.99$.

The mass distribution for the events passing the muon pair selection is shown 
in figure \ref{Fig:Mass}b. 
A clear peak over a flat background is observed.
The signal region, $2.95 < M_{\mu^+\mu^-} < 3.25$~GeV, contains 289 events.
The cross sections and angular distributions presented below are obtained
by calculating acceptances and background contributions for this range.
An unbinned likelihood fit to the sum of a Gaussian signal plus a flat
background gives a value of $3.086 \pm 0.003$~GeV for the mass,
$38 \pm 3$~MeV for the rms width and $266 \pm 17$ for the number of
events attributable to $J/\psi$ production in the mass range $2 <
M_{\mu^+\mu^-} < 4$~GeV. 

\section{\bf Monte Carlo Simulation and Acceptance Calculation}
\label{Sect:MCAcc}

The reaction $e^+~p \rightarrow e^+~J/\psi~p$ (figure \ref{Fig:Feynman}a)
was modelled using the DIPSI Monte Carlo program \cite{Ref:DIPSI}.
This Monte Carlo is based on the model of Ryskin \cite{Ref:Ryskin} in
which it is assumed that the exchanged photon fluctuates into a
$c\bar{c}$ pair which then interacts with a gluon ladder emitted by
the incident proton.
The events are generated with a cross section proportional to $W^{\delta}$
and with an exponential $t$ distribution proportional to $exp(-b|t|)$.
Good agreement between the generated and observed distributions is
obtained for $\delta=1$ and $b=4$~GeV$^{-2}$.
In order to determine the systematic error on the acceptance $\delta$
was varied in the range $0 < \delta < 2$.
The acceptance was found to be insensitive to the variation of $b$ in
the range $3 < b < 5$~GeV$^{-2}$.

Events were generated in the $W$ range $20<W<210$~GeV and between 
$Q^2_{min}$ and $\qsq = 4$~GeV$^2$.
The centre of mass decay of the \jpsi~was generated with a 
$(1+\alpha \cos^2 \theta_h)$ distribution with $\alpha = 1$.
Varying the value of $\alpha$ from 1 to 0.4, corresponding to about one
standard deviation variation around the measurement presented in 
section \ref{Sect:Decay}, the acceptance grows by less than $10$\%.
A systematic error due to this uncertainty is included in the total
systematic error as described in section \ref{Sect:Systematics}.
The effects of positron initial and final state radiation and that of
vacuum polarisation loops were neglected; the effects on the
integrated cross section have been estimated to be smaller than 4\%
\cite{Ref:ZEUSRho}.

The events were then passed through a detailed simulation of the ZEUS 
detector and trigger.
Parameterisations of noise distributions obtained from data taken with
a random trigger were used to simulate the calorimeter noise
contribution to the energy measurements.
The simulated events were subjected to the same reconstruction and 
analysis programs as the data.
The distributions of the reconstructed kinematic quantities obtained using
DIPSI are in good agreement with those from the data.
The overall acceptance was obtained as the ratio of the number of accepted
Monte Carlo events to the number generated in the selected kinematic range.
The acceptance, calculated in this manner, accounts for the geometric
acceptance, for the detector, trigger and reconstruction efficiencies, and 
for the detector resolution.
Table \ref{Tab:GammaP} shows the acceptances in various $W$ ranges determined
for each decay mode.

\section{\bf Background}
\label{Sect:Background}

In addition to elastic \jpsi~photoproduction, the following processes may 
contribute to the final sample:
{\parindent 0pt
\begin{itemize}
    \item The Bethe-Heitler process in which a lepton pair is produced by the
         fusion of a photon radiated by the positron with a photon radiated
         by the proton.
         This process was simulated using the LPAIR Monte Carlo
         \cite{Ref:LPair} which was used to generate events in which the
         proton remains intact (`elastic' events) and events in which
         the proton dissociates (`dissociative' events).
         The size of the Bethe-Heitler contribution to the non-resonant
         background is shown in figure \ref{Fig:Mass} where the 
         $\ell^+\ell^-$~mass distributions are plotted.
         The QED cross section \cite{Ref:Vermaseren} for the elastic and 
         dissociative Bethe-Heitler processes have been used to determine 
         the normalisation of the appropriate LPAIR Monte Carlo sample.
         Figure \ref{Fig:Mass} shows that the Bethe-Heitler process saturates
         the non-resonant background in the muon channel and is the dominant
         source of non-resonant background in the electron channel.
         The calculated background due to the Bethe-Heitler process in the
         signal region is $38\pm1$ for the electron channel and
         $23\pm1$ for the muon channel.
    \item Pions misidentified as electrons in the electron sample.
         For $e^+e^-$ masses larger than 2.5~GeV the Bethe-Heitler
         contribution saturates the non-resonant background.
         The residual contribution of misidentified pions in the final sample
         was shown to be less than 1.5\% by studying the distribution of 
         $dE/dX$ obtained using the pulse height information from the CTD.
         No subtraction has been made for pion misidentification. 
         A systematic error of $-1.5$\% attributed to the uncertainty in the
         pion contamination was included in the final systematic error.
    \item \jpsi~produced via the production and decay of $\psi'$.
         The only $\psi'$ decay mode giving a significant contribution to the
         \jpsi~signal is $\psi' \rightarrow \jpsi \pi^0 \pi^0$.
    \item Proton dissociative \jpsi~production (figure \ref{Fig:Feynman}b).
         The EPSOFT Monte Carlo was used to simulate this process.
         EPSOFT is based on the assumption that the diffractive cross
         section is of the form 
         $d\sigma / d|t| dM_N^2\propto {e^{-b_d |t|}}/{M_N^{\beta}}$
         where $M_N$ is the mass of the dissociative system.
         The simulation of the hadronisation of the dissociative system
         includes a parameterisation of the resonance spectrum.
         To cross-check the results the generator PYTHIA \cite{Ref:PYTHIA} 
         was also used which contains a different parameterisation of the
         resonance spectrum. 
\end{itemize}
}

After the subtraction of the Bethe-Heitler contribution, the production of
$J/\psi$~mesons via the decay of the $\psi'$ and proton 
dissociative \jpsi~production are the only significant sources of
background and will be discussed separately below.

The $\psi'$ contribution was determined using a sample of events in 
which the $\psi'$ decayed to a muon pair (branching ratio 
$\BR_1 = (0.77\pm0.17)\%$ \cite{Ref:PDG}).
This sample was obtained using the same cuts as those used to isolate the 
$J/\psi \rightarrow \mu^+ \mu^-$ sample (see section \ref{Sect:Event}).
A signal of $N_1 = 7 \pm 4$ events was found at the $\psi'$ mass in a
sample for which the integrated luminosity, $\Lumi_1$, was 
$2.70\pm0.04$~pb$^{-1}$. 
The corresponding acceptance, $\Acce_1$, computed with DIPSI, was
$\Acce_1=0.35$.
The number of events from $\psi'$ production entering the elastic 
$\jpsi \rightarrow \mu^{+}\mu^{-}$~
sample via the decay $\psi' \rightarrow J/\psi \pi^0 \pi^0$ was
estimated using the formula
\begin{equation}
    N_C = \frac{N_1}{\Acce_1 ~\Lumi_1 ~\BR_1} \Acce^{\mu}_C \Lumi_C \BR_C \BR,
\label{Eq:PsiPrime}     
\end{equation}
where $\BR=(6.01\pm0.19)\%$ is the branching ratio for the decay
$J/\psi \rightarrow \mu^+\mu^-$, $\BR_C$ is that for the decay
$\psi^{'} \rightarrow \psi \pi^0\pi^0$ 
($\BR_C=(18.4 \pm 2.7)\%)$ \cite{Ref:PDG}, $\Lumi_C$ 
is the luminosity from which the muon sample defined in section
\ref{Sect:Event} was drawn ($\Lumi_C = 1.9$~pb$^{-1}$) and
$\Acce^{\mu}_C$ is the acceptance for the process  
$e^+ p \rightarrow e^+ \psi' p$ ($\psi'\rightarrow\mu^+\mu^-\pi^0\pi^0$),
using DIPSI $\Acce^{\mu}_C = 0.28$. 
The formula (\ref{Eq:PsiPrime}) leads to a $\psi'$ contamination of
($2.3\pm1.4$)\%. 
This result was cross-checked by selecting events in which the $\psi'$ decayed
into $\mu^+\mu^-\pi^+\pi^-$.
In this case $7\pm3$ events were found at the $\psi'$ mass and a contamination
of $(3.4\pm1.4)$\% was estimated.
The two results may be combined to give a final estimate of the $\psi'$ 
contamination of $(3\pm1)$\%.
This contamination was subtracted from both the electron and muon sample.

The proton dissociative process is characterised by a cross section of
the form 
\begin{equation}
    \frac{d\sigma}{d|t|dM_N^2} \propto \frac{e^{-b_{d}|t|}}{M_N^{\beta}} .
\end{equation}
In order to estimate the value of $b_{d}$, dissociative events were
selected in which the \jpsi~was accompanied by an energy deposit in
the inner ring of FCAL or in the PRT.
The value $b_d=1$~GeV$^{-2}$ was found to give the best
description of the $p_{T J/\psi}$ distribution of the PRT tagged sample.
The systematic error in the dissociative contribution caused by the uncertainty
in $b_d$ was estimated by varying $b_d$ in the range 
$0.4 < b_d < 2$~GeV$^{-2}$. 
This assumption is consistent with the result $b_d=1.6\pm0.3\pm0.1$~GeV$^{-2}$
reported by the H1 collaboration\cite{Ref:H1JPsi94}.
The value $\beta=2.25$ was used as the central value in the simulation
of the $M_N$ distribution and $\beta$ varied in the range $2 < \beta <
2.5$ to estimate the systematic error.
This assumption is consistent with the result $\beta = 2.20 \pm 0.03$
recently obtained at Fermilab for the diffractive dissociation of the
proton in $\bar{p}p$ collisions \cite{Ref:FermiLab}. 
The mass of the nucleonic system was generated in the range 
$(1.25~{\rm GeV}^2) \leq M_N^2 \leq 0.1~W^2$. 

The proton dissociative contribution to the electron sample was determined by
selecting a sample, $D_e$, for which the requirement that 
$E_{Tot} - E_{J/\psi} < 1$~GeV was replaced by the three cuts
$E_{F} > 1 {\rm~GeV,~} E_{B} < 1 {\rm~GeV~and~} E_{R} < 1 {\rm~GeV}$.
$E_F$, $E_B$ and $E_R$ were calculated by summing the energy in the FCAL,
BCAL and RCAL respectively.
The calorimeter cells associated with the electron candidates were excluded
from these sums.
The cut on $E_F$ selects dissociative events in which energy is deposited in 
the proton direction, while the cut on $E_R$ ensures that events in which 
the scattered positron is detected in RCAL do not enter the sample.
The cut on $E_B$ ensures that inelastic events depositing energy in
BCAL also do not enter the sample.
The proton dissociative sample, $D_e$, was further examined by studying the 
distribution of the energy weighted pseudorapidity defined by
\begin{equation}
\bar{\eta}_{C}=\frac{\Sigma_i E_{i} \eta_{i} }{\Sigma_i E_{i} } ,
\end{equation}
where $E_{i}$ is the energy of a calorimeter cell and $\eta_{i}$ is 
the pseudorapidity of the cell and the sum runs over all cells containing 
more than 200~MeV but excluding those matched to the tracks forming the 
\jpsi~candidate.
The distribution of $\bar{\eta}_{C}$ for dissociative events, simulated using 
the EPSOFT Monte Carlo, is strongly peaked at $\bar{\eta}_{C} > 2$.
In the sample $D_e$ there are 2 events for which
$\bar{\eta}_{C} > 2$.
The ratio of the number of EPSOFT events passing the elastic cuts to
the number with $E_F>1$~GeV, $E_B<1$~GeV, $E_R<1$~GeV and
$\bar{\eta}_C>2$ was 58. 
This leads to a dissociative contribution to the elastic \jpsi~to electron 
sample of $(33^{+43+7+~0}_{-12-6-18}){\rm\%}$.
The first error is statistical and the second error is the systematic error 
resulting from the allowed variation of $\beta$ in the Monte Carlo generation
of dissociative events.
When the calculation is repeated with EPSOFT replaced by PYTHIA the result
differs by -18\% from that reported above.
The third error quoted in the dissociative contribution reflects this
uncertainty in the simulation of the dissociative final state.
The change in the dissociative contribution obtained when $b_d$ was
varied in the range $0.4<b_d<2$~GeV$^{-2}$ was found to be negligible.

The same procedure was applied to the muon sample with the only
difference that the cut on $\bar{\eta}_C$ was not applied.
The proton dissociative sample obtained contained 7 events and
the ratio of the number of EPSOFT events passing the elastic cuts 
to the number with $E_F>1$~GeV, $E_B<1$~GeV, $E_R<1$~GeV was 11. 
This leads to a dissociative contribution of 
$(29 \pm 11 ^{+6~+~0}_{-5~-10}){\rm \%}$.

Independent estimates of the dissociative contribution were made 
using dissociative events tagged by the PRT.
EPSOFT was used to estimate the fraction of untagged dissociative events
in the elastic sample since it was found that PYTHIA gives a poor
description of the multiplicity distribution observed in the PRT.
The dissociative contamination estimated in this way was $(34 \pm 8){\rm\%}$
for the electron channel and $(27 \pm 8){\rm\%}$ for the muon channel.
The errors quoted are statistical only.

The four independent results were combined to give a final estimate
of the dissociative contribution of $(30 \pm 5 ^{+7~+~0}_{-6~-10})$\%.

\section{\bf Systematic Errors}
\label{Sect:Systematics}

Several factors contribute to the systematic errors in the 
elastic J/$\psi$ cross section measurement. In the following they are 
divided in two categories: 
{\it decay channel specific errors} and 
{\it common systematic errors}. The first category contains
systematic errors specific to the electron or muon decay channel,
while the second contains systematic errors common to both decay
channels. 
Table \ref{Tab:SysErr} summarises all these systematic errors.

\paragraph
{\parindent 0pt
{\bf Decay channel specific errors}:
}
\begin{itemize}
\item {\it Trigger}: 
For the electron channel, the dominant systematic error due to the
FLT acceptance is given by the requirement $E_{Tot} > 2$ GeV.
At the SLT the dominant systematic error is contributed by the
simulation of the calorimeter noise.
For the muon channel, the dominant systematic error is contributed
by the uncertainties in the simulation of the trigger threshold and
the CTD-FLT track reconstruction.  
No systematic error in either channel is attributed to the TLT acceptance
since all cuts are superseded by more stringent requirements offline.

\item {\it Event selection}: 
In this class we include the systematic errors due to uncertainties in
the measurement of momentum, transverse momentum, $|\eta|$ and the
choice of the mass window. 
For the electron channel uncertainties in the cuts used to define an
electron cluster also contribute.
For the muon channel this class also contains the uncertainties coming
from the collinearity cut.
Each cut was varied within a range determined by the resolution of the
quantity in question and the changes induced in the results were taken
as an estimate of the corresponding systematic error.
The different systematic errors were summed in quadrature.

\item {\it Pion misidentification}: 
This class applies to the electron channel only; the method used to
determine the systematic error was described in section
\ref{Sect:Background}.

\item {\it Muon chamber efficiency}: 
The systematic error attributed to errors in the muon chamber
reconstruction efficiency was estimated by using cosmic ray events.

\item {\it Branching ratio}: 
The error on the branching ratio $J/\psi \rightarrow \ell^+ \ell^-$ as
quoted in \cite{Ref:PDG}. 
\end{itemize}

{\parindent 0pt
{\bf Common systematic errors}:
}
\begin{itemize}

\item {\it Acceptance}: 
The uncertainty in the acceptance was estimated by varying the
parameters $b$ and $\delta$ as described in section \ref{Sect:MCAcc}. 

\item {\it Elastic definition}: 
The systematic uncertainty contributed by the criterion used to
classify an event as elastic was estimated by changing the elastic
definition: $E_{Tot}-E_{J/\psi} < 1 $ GeV to $E_{Tot}-E_{J/\psi} < 0.7$ GeV
and to $E_{Tot}-E_{J/\psi} < 1.3$ GeV.

\item {\it Radiative corrections}: 
The effects of positron initial and final state radiation and that of
vacuum polarisation loops were neglected; the effects on the
integrated cross section have been estimated to be smaller than 4\%
\cite{Ref:ZEUSRho}.
We take 4\% as an estimate of the systematic error attributable
to this source. 

\item {\it Helicity distribution:}
The centre of mass decay of the \jpsi~was generated with a 
$(1+\alpha \cos^2 \theta_h)$ distribution.
The systematic error was evaluated by varying the value of $\alpha$
from 1 to 0.4. 

\item {\it $M_N$ distribution in proton dissociation}: 
As explained in section \ref{Sect:Background} this is obtained by changing
the parameter $\beta$ in the range $2\leq \beta \leq 2.5$.

\item {\it Model of dissociation:}
The dependence on the modelling of the hadronic final state in proton
dissociation was obtained by comparing the contamination obtained
using PYTHIA with that obtained using EPSOFT (see section
\ref{Sect:Background}).

\item {\it $\psi^{'}$ contamination}:
As explained in section \ref{Sect:Background} the systematic error on 
the $\psi'$ contribution is 1\%.

\item {\it Luminosity}:
As indicated in section \ref{Sect:ZEUS} the uncertainty of the
luminosity determination is 1.5\%.
\end{itemize}

\section{\bf Results}
\label{Sect:Results}

\subsection{\bf Integrated Cross Sections}
\label{Sect:Xsection}

The cross section for elastic \jpsi~electroproduction is given by
\begin{equation}
    \sigma_{ep \rightarrow e J/\psi p} = \frac{N_{Evt}}{\Lumi \Acce \BR},
\label{Eq:EPXSect}
\end{equation}
where \Lumi~is the integrated luminosity, \Acce~is the acceptance, \BR~is the 
branching ratio for \jpsi~to decay into electron or muon pairs \cite{Ref:PDG} 
and $N_{Evt}$ is the number of signal events after background subtraction.
$N_{Evt}$ and $\Acce$ were determined in the signal regions defined for
the electron and muon channels in section \ref{Sect:Event}.
In the range $40 < W < 140$~GeV and for $Q^2_{min} < Q^2 < 4$~GeV$^2$ the
\jpsi~electroproduction cross section is
\begin{equation}
    \sigma_{ep \rightarrow e J/\psi p} = 5.37 \pm 0.30 {\rm (stat.)} 
                                      ^{+0.69}_{-0.86} {\rm (syst.)}
                                       ^{+0.54}_{-0} {\rm (model)}{\rm~nb,}
\end{equation}
using the electron sample and
\begin{equation}
    \sigma_{ep \rightarrow eJ/\psi p} = 5.04 \pm 0.32 {\rm (stat.)} 
                                      ^{+0.62}_{-0.78} {\rm (syst.)}
                                       ^{+0.50}_{-0} {\rm (model)}{\rm~nb,}
\end{equation}
using the muon sample.
The model error quoted above is due to the difference between the
value of the dissociative contribution estimated using EPSOFT and 
using PYTHIA.
In the systematic error we have summed in quadrature all the 
decay-channel-specific errors and the common systematic errors.
The electron and muon cross section results are compatible with each other and
with previous measurements in the same $W$ range
\cite{Ref:ZeusJPsi93, Ref:H1JPsi93, Ref:H1JPsi94}.

\subsection{Photoproduction Cross Section}
\label{Sect:GammaP}

\begin{figure}[p]
    \vspace*{-0.8cm}
\begin{center}
\epsfig{file=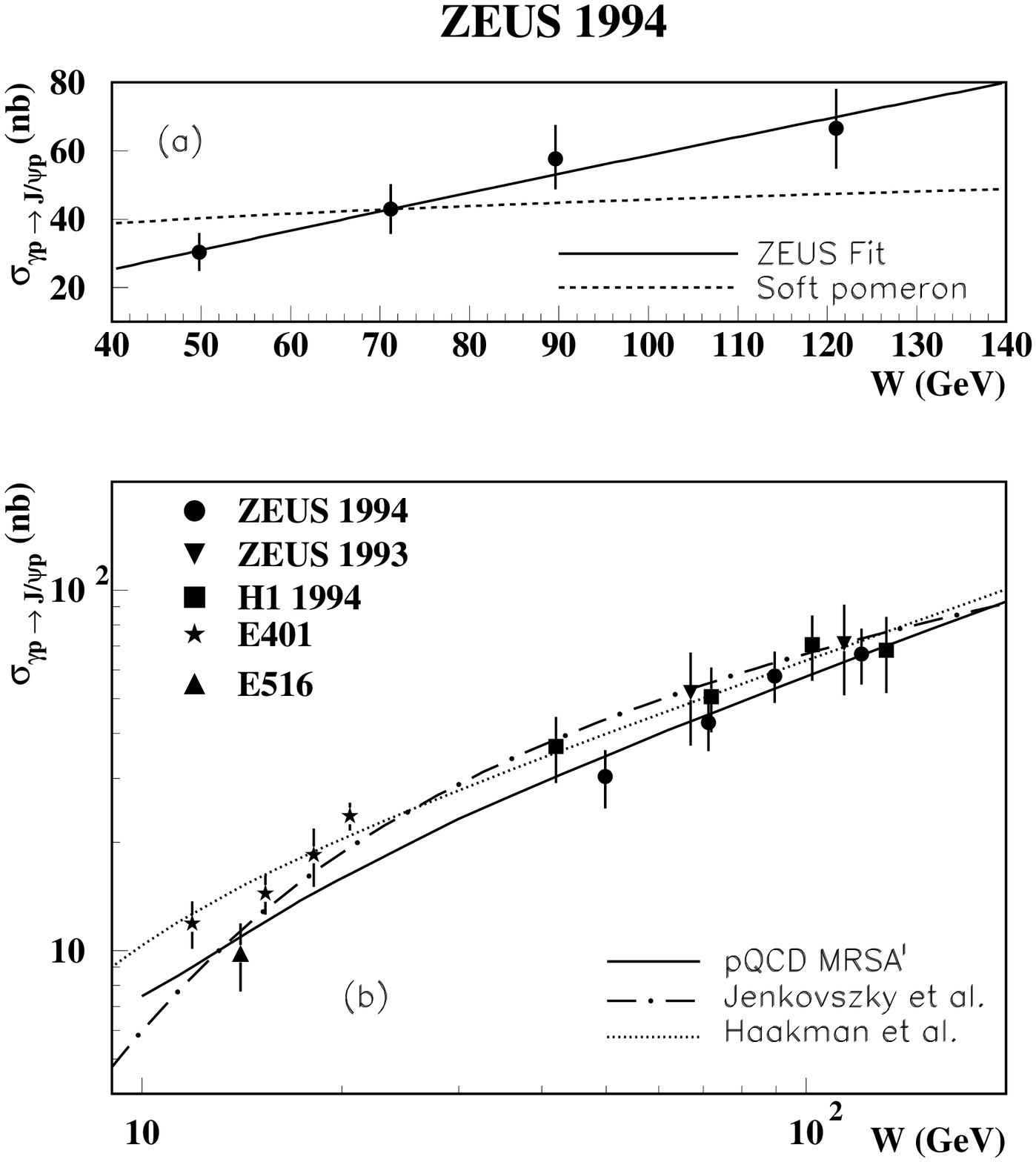,%
      width=16cm%
        }
    \vspace*{-2.0cm}
    \caption{
             The elastic \jpsi~photoproduction cross section as a function of 
             $W$.
             (a) Shows the results of this analysis. 
             The error bars represent the statistical and systematic errors 
             added in quadrature.
             The solid line shows the result of the fit to the data
             using the expression $\sigma_{\gamma p \rightarrow J/\psi p} \propto W^{\delta}$.
             As described in the text the value $\delta = 0.92 \pm 0.14 \pm 0.10$ was obtained.
             The dashed line shows the prediction of a soft pomeron model
             \protect\cite{Ref:DL} in which $\delta \approx 0.22$.
             (b) The results of this analysis (solid circles)
             are compared to data from H1, ZEUS and the results of
             lower energy measurements \protect\cite{Ref:FixedTrgtJPsi, Ref:E401}.
             The result of a pQCD calculation \protect\cite{Ref:RyskinRoberts}
             in which the MRS-A$^{\prime}$ \protect\cite{Ref:MRSG} parton 
             distributions have been used is shown as the solid line.
             The result of the calculation presented in \protect\cite{Ref:Kaidalov}
             is shown by the dotted line.
	     The result of the calculation presented in \protect\cite{Ref:Paccanoni}
             is shown by the long dash dotted line.
             }
\label{Fig:GammaP}
\end{center}
\end{figure}

The photoproduction cross section is related to the $ep$ cross section by
\cite{Ref:Flux}
\begin{equation}
    \sigma_{\gamma p \rightarrow J/\psi p} =
     \frac{\int \Phi(y, Q^2) \sigma_{\gamma p \rightarrow J/\psi p}
                                              \left(y, Q^2 \right) dy dQ^2}
          {\int \Phi(y, Q^2) dy dQ^2}            =
     \frac{\sigma_{ep \rightarrow e J/\psi p}}{\Phi_T} ,
\label{Eq:GammaPXSect}
\end{equation}
where $\sigma_{\gamma p \rightarrow J/\psi p}$ is the mean cross section
in a range of $W$ and $\Phi_T$ is the effective flux of virtual photons 
accompanying the positron.
The integrals run over the full range of $Q^2$ and from $y_{min}=W^2_{min}/s$
to $y_{max}=W^2_{max}/s$ where $W_{min}$ and $W_{max}$ are the minimum and
maximum values of $W$ respectively.
The photoproduction cross section has been determined in four $W$ bins.
The results for each of the lepton decay modes and the combined results are 
reported in table \ref{Tab:GammaP}.
The procedure described in section \ref{Sect:Xsection} was used to
calculate the errors on the cross sections presented in table
\ref{Tab:GammaP}. 
For the combined results the following procedure was used.
The weighted mean cross section was calculated; the weights being
obtained by summing the statistical and decay channel specific errors  
in quadrature. 
The first error reported on the combined results in table
\ref{Tab:GammaP} is the error on the weighted mean, the
second error is the sum of the common systematic errors added in
quadrature. 
The third error reported on the combined results in table
\ref{Tab:GammaP} is the systematic error associated with the model of
diffraction. 
The combined results are shown in figure \ref{Fig:GammaP} where 
$\sigma_{\gamma p \rightarrow J/\psi p}$ is plotted as a function of 
$W$.
The points are plotted at the mean values of $W$ reported in table
\ref{Tab:GammaP}.
A clear growth of $\sigma_{\gamma p \rightarrow J/\psi p}$ with $W$ is 
observed over the $W$ range covered by this experiment.

The ZEUS data in the range $40<W<140$~GeV were fit to the form 
$\sigma_{\gamma p \rightarrow J/\psi p} \propto W^{\delta}$ with
the result $\delta = 0.92 \pm 0.14 {\rm~(stat.)} \pm 0.10 {\rm~(syst.)}$.
The systematic error was obtained as follows.
For each source of systematic error in turn the cross sections were
displaced from their central values, the fit was performed and the
value $\delta_{si}$ recorded.
The systematic error on $\delta$ was taken to be 
$\sqrt{\sum_i \left( \delta - \delta_{si} \right)^2 }$.
The result of the fit is shown in figure \ref{Fig:GammaP}a.
This value of $\delta$ disfavours
that expected in the Donnachie-Landshoff model \cite{Ref:DL}
(the soft pomeron model) in which $\delta$ is expected to take the value
$\delta=0.22$ in this $W$ range.
The curve corresponding to the soft pomeron model is shown in figure
3a as a dotted line arbitrarily normalised to the second ZEUS data point.

It is interesting to compare the ratio, $R(\frac{J/\psi}{\rho})$, of
the cross section for elastic \jpsi~photoproduction to the cross
section for elastic $\rho$ production as a function of $W$.
At $W \simeq 12$~GeV 
$R(\frac{J/\psi}{\rho})=\left( 1.21\pm0.20 \right) \times 10^{-3}$
while at $W\simeq15$~GeV 
$R(\frac{J/\psi}{\rho})=\left( 1.67\pm0.23 \right) \times 10^{-3}$
\cite{Ref:E401, Ref:E516, Ref:Egloff}.
The results presented in the present paper may be combined with those
presented in reference \cite{Ref:ZEUSRho} to determine that
$R(\frac{J/\psi}{\rho})=\left( 2.94\pm0.74 \right) \times 10^{-3}$
at $W \simeq 70$~GeV showing that
$R(\frac{J/\psi}{\rho})$ rises with $W$.
These values are to be compared with $R(\frac{J/\psi}{\rho})=\frac{8}{9}$ 
expected on the basis of the quark charges and a flavour independent
production mechanism.

The data are replotted in figure \ref{Fig:GammaP}b together with other
measurements of elastic \jpsi~photoproduction.
The results of two pomeron models \cite{Ref:Kaidalov, Ref:Paccanoni}
are shown in figure \ref{Fig:GammaP}b. 
In the model of reference \cite{Ref:Kaidalov} the effective pomeron
intercept is assumed to depend upon $\bar{Q}^{2}_{HKK}=c M^{2}_{c}+Q^2$,
where $M_c$ is the mass of the charm quark and the constant $c \approx 1$.
The model of reference \cite{Ref:Paccanoni} assumes a fixed pomeron
intercept but includes both a scale dependent pomeron coupling and a
mass threshold function.
Both models give a good description of the data.

Attempts have been made to describe elastic \jpsi~production in
perturbative QCD, pQCD.
In the approach of Ryskin \cite{Ref:Ryskin}
the pomeron is described as a gluon ladder evaluated in the leading logarithm
approximation.
In this model the cross section is proportional to 
$\left[ \alpha_s \bar{x} g(\bar{x}, \bar{q}^2) \right]^2$,
where $\alpha_s$ is the strong coupling constant (assumed fixed and
set equal to 0.25) and $\bar{x} g(\bar{x}, \bar{q}^2)$ is the gluon
momentum density in the proton.
The quantities $\bar{x}$ and $\bar{q}^2$ are given by
\begin{equation}
    \bar{x}   = \frac{\qsq + M^2_{J/\psi} - t}{W^2}        \\
    \bar{q}^2 = \frac{\qsq + M^2_{J/\psi} - t}{4}
\end{equation}
and give the effective momentum fraction and scale at which the gluon density 
is probed respectively.
In the present case both $Q^2$ and $|t|$ are negligible in comparison
to $M^2_{J/\psi}$.
For elastic \jpsi~photoproduction $\bar{q}^2$ takes a value of approximately 
2.5~GeV$^2$ \cite{Ref:Ryskin} while the measurements presented here
are sensitive to values of $\bar{x}$ in the range 
$0.4 \times 10^{-3} < \bar{x} < 6 \times 10^{-3}$ \cite{Ref:ZeusJPsi93}.
If a gluon distribution of the form 
$\bar{x} g\left( \bar{x}, {Q}^2 \right) \propto \bar{x}^{-\lambda}$
is assumed then the $W$ dependence of 
$\sigma_{\gamma p \rightarrow J/\psi p}$ may be written
$\sigma_{\gamma p \rightarrow J/\psi p} \propto W^{4 \lambda}$.
The value of $\delta$ reported above gives 
$\lambda = 0.23 \pm 0.04 \pm 0.03$.
This is consistent with our measurement of
the gluon distributions based on an analysis of the scaling violations
of $F_2$ extrapolated back to $Q^2 = 2.5$~GeV$^2$ \cite{Ref:ZEUSGluon}.

Figure \ref{Fig:GammaP}b shows the results of the pQCD calculation of
$\sigma_{\gamma p \rightarrow J/\psi p}$ 
presented in \cite{Ref:RyskinRoberts} which extends the Ryskin model beyond
leading order and includes the effects of the relativistic motion of
the $c$ and $\bar{c}$ within the \jpsi~and the rescattering of the
$c\bar{c}$ pair on the proton.
Good agreement with the data is obtained using the MRS-A$^{\prime}$
\cite{Ref:MRSG} parton distributions.
Other choices of parton distributions compatible with HERA
measurements of $F_2$ also give an acceptable description of the $W$
dependence of $\sigma_{\gamma p \rightarrow J/\psi p}$ over the range 
$40 < W < 140$ GeV.

\subsection{\bf Differential Cross Sections}
\label{Sect:t_Dist}

\begin{figure}[p]
\begin{center}
\epsfig{file=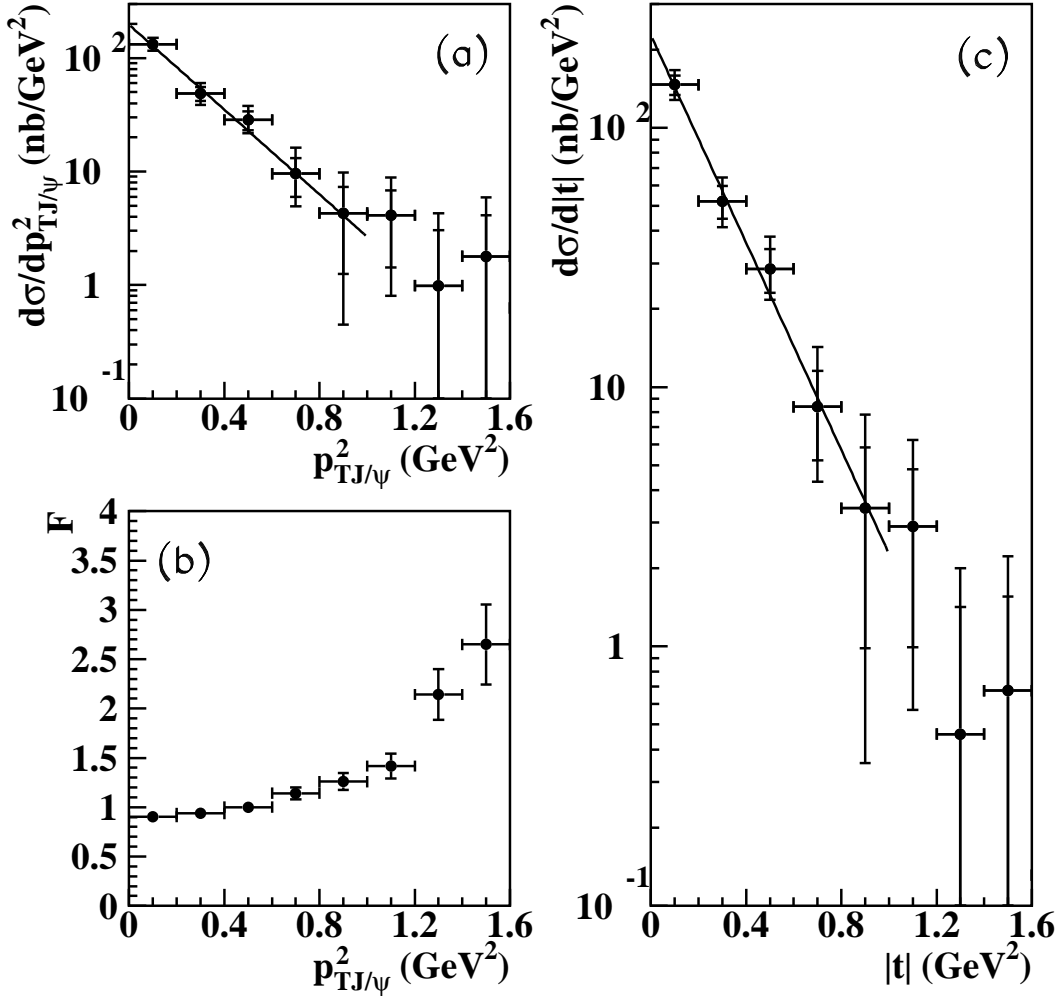,%
      width=15cm%
        }
    \caption{The distribution of transverse momentum squared for \jpsi~produced in the 
             reaction $\gamma p \rightarrow J/\psi p$ in the kinematic range 
             $40 < W < 140$~GeV. 
             (a) The differential cross section 
             ${d\sigma}/{dp_{T J/\psi}^2}$.
             The data are shown as the points and the result of the 
             exponential fit in the range $p_{T}^2<1$~GeV$^2$ is shown 
             as the solid line.
             (b) The correction factor, $F$, required to obtain the $|t|$
             distribution from the $p^{2}_{T J/\psi}$ distribution by accounting
             for the $Q^2$ of the photon.
             (c) The differential cross section ${d\sigma}/{d|t|}$. 
             The result of the exponential fit in the range
             $|t|<1$~GeV$^2$ is shown as the solid line. In (a) and
             (c) the inner error bars represent the statistical and
             decay-channel-specific errors added in quadrature,
             the outer ones statistical, decay-channel-specific
             errors and common systematic errors added in quadrature. 
             }
\label{Fig:pt2_t}
\end{center}
\end{figure}

Figure \ref{Fig:pt2_t}a shows the differential photoproduction cross
section ${d \sigma}/{dp_{T J/\psi}^2}$ for the full $W$ range 
($40 < W < 140$~GeV).
The results from the electron and  
muon samples have been combined using the procedure described 
in section \ref{Sect:GammaP}.
The contribution from proton dissociative \jpsi~production and the 
Bethe-Heitler process have been subtracted bin by bin.
The cross section exhibits the exponential fall characteristic of
diffractive processes.
A binned likelihood fit to the form
\begin{equation}
    \frac{d \sigma}{dp_{T J/\psi}^2} = A e^{-b_{p_T} p_{T J/\psi}^2}
\label{Eq:dsdpt2}
\end{equation}
was performed in which the function in equation \ref{Eq:dsdpt2} was integrated
and compared with the measured cross section bin by bin.
Fitting over the range $ p_{T J/\psi}^2 < 1$~GeV$^2$ gives the result
\begin{equation}
    b_{p_T} = 4.3 \pm 0.4 ^{+0.4}_{-0.6} {\rm ~GeV}^{-2}.
\end{equation}

The differential cross section ${d \sigma}/{d|t|}$ may be obtained
by dividing ${d\sigma}/{dp_{T J/\psi}^2}$ bin by bin by a factor which
corrects for the small $Q^2$ of the photon.
Figure \ref{Fig:pt2_t}b shows the correction factor, $F$, which is
slowly varying and close to 1 for $|t|<1$~GeV$^2$.
The differential cross section ${d \sigma}/{d|t|}$ obtained in this way is
plotted in figure \ref{Fig:pt2_t}c.
Again, the cross section exhibits an exponential fall and
a binned likelihood fit to the form
\begin{equation}
    \frac{d \sigma}{d|t|} = A e^{-b|t|}
\label{Eq:dsdt}
\end{equation}
was performed in which the function in equation \ref{Eq:dsdt} was integrated
and compared with the measured cross section bin by bin.
Fitting over the range $|t| < 1$~GeV$^{2}$ gives the result
\begin{equation}
    b = 4.6 \pm 0.4 ^{+0.4}_{-0.6} {\rm~GeV}^{-2} .
\label{Eq:bresult}
\end{equation}
The systematic error contains the contribution coming from the
uncertainty in the correction factor $F$.
The fit for $b$ was repeated for $|t|<0.8$~GeV$^{2}$ and
$|t|<1.2$~GeV$^{2}$.
The small changes in $b$ obtained are included in the systematic error
quoted in equation \ref{Eq:bresult}.
The size of the statistical and systematic errors on the parameter
$b$ prevents us from investigating the dependence of $b$ on $W$ using
the data presented here. 
The slope is in agreement with the result obtained by the H1
collaboration \cite{Ref:H1JPsi93, Ref:H1JPsi94} in the same $W$ range.
We have previously determined the parameter $b$ in elastic $\rho$, $\omega$
and $\phi$ photoproduction to be 
$9.8\pm0.8{\rm~stat.}\pm1.1{\rm~syst.}$~GeV$^{-2}$ \cite{Ref:ZEUSRho},
$10.0\pm1.2{\rm~stat.}\pm1.3{\rm~syst.}$~GeV$^{-2}$ \cite{Ref:ZEUSOme} and
$7.3\pm1.0{\rm~stat.}\pm0.8{\rm~syst.}$~GeV$^{-2}$ \cite{Ref:ZEUSPhi}
respectively.
In geometrical models of vector meson production these results may be
interpreted as indicating that the radius of the \jpsi~is smaller than
that of the $\rho$, $\omega$ and $\phi$.
When the parameter $b$ is measured in exclusive $\rho$ production in
deep inelastic scattering for $Q^2$ values in the range 
$7 \sleq Q^2 \sleq 25$~GeV$^2$ a value of
$5.1^{+1.2}_{-0.9}\pm1$~GeV$^{-2}$ is obtained which is significantly
smaller than the slope obtained in elastic $\rho$ photoproduction
\cite{Ref:ZEUSDISRho}.
Thus, in exclusive $\rho$ production $b$
falls as $Q^2$ is raised from 0 reaching a value of 
$5.1^{+1.2}_-{-0.9}\pm1.0$~GeV$^{-2}$ at $Q^2$ of order
10~GeV$^{2}$ comparable to that reported here for
$J/\psi$~photoproduction where
the hard scale in the scattering process may be set by $M_{J/\psi}^2$.

\subsection{\bf Decay Angular Distributions}
\label{Sect:Decay}

The $\jpsi$ decay angular distributions can be used to determine
elements of the $J/\psi$ spin-density matrix~\cite{Ref:RhoPhiAng}. 
In the $s$-channel helicity frame the \jpsi~is at rest and the quantisation 
axis is taken to lie along the \jpsi~direction in the photon-proton centre of 
mass system. 
The decay angular distribution is a function of $\theta_h$ and $\phi_h$, the
polar and azimuthal angles of the positive lepton in the helicity frame.
The angular distributions can be shown to be \cite{Ref:Helicity}
\begin{equation}
  \frac{1}{N} \frac{dN}{d \cos \theta_h} = \frac{3}{8} \left[
                               1 + r^{04}_{00} + 
                 \left( 1 - 3 r^{04}_{00} \right) \cos^2 \theta_h
                                       \right],
\end{equation}
\begin{equation}
  \frac{1}{N} \frac{dN}{d \phi_h} = \frac{1}{2 \pi} \left[
                               1 + r^{04}_{1-1} \cos 2 \phi_h 
                                       \right].
\end{equation}
In the present experiment $\qsq \approx 0$~GeV$^2$ so that the \jpsi~is expected to be
produced predominantly by transverse photons. 
If $s$-channel helicity is conserved (SCHC) then the spin density matrix 
parameters $r^{04}_{00}$ and $r^{04}_{1-1}$ should be close to zero.
Under the assumption of SCHC, $r^{04}_{00}$ can be related to the ratio of the
photoproduction cross sections for longitudinal and transverse photons
\begin{equation}
        R = \frac{1}{\epsilon} \frac{r^{04}_{00}}{1-r^{04}_{00}}      \\
 \epsilon = \frac{2 \left( 1-y \right) }{1+ \left( 1-y \right)^2 - 2 \left( 1-y \right)
                     \frac{Q^2_{min}}{Q^2} }
     ,
\label{Eq:RDef}
\end{equation}
where $\epsilon$ is the virtual photon polarisation, i.e. the ratio of
the flux of longitudinally polarised photons to the flux of
transversely polarised photons.
The mean value of $\epsilon$ over the kinematic range of $Q^2$ and $y$
sampled by the present experiment is $\epsilon = 1.043$.

\begin{figure}[p]
\begin{center}
\epsfig{file=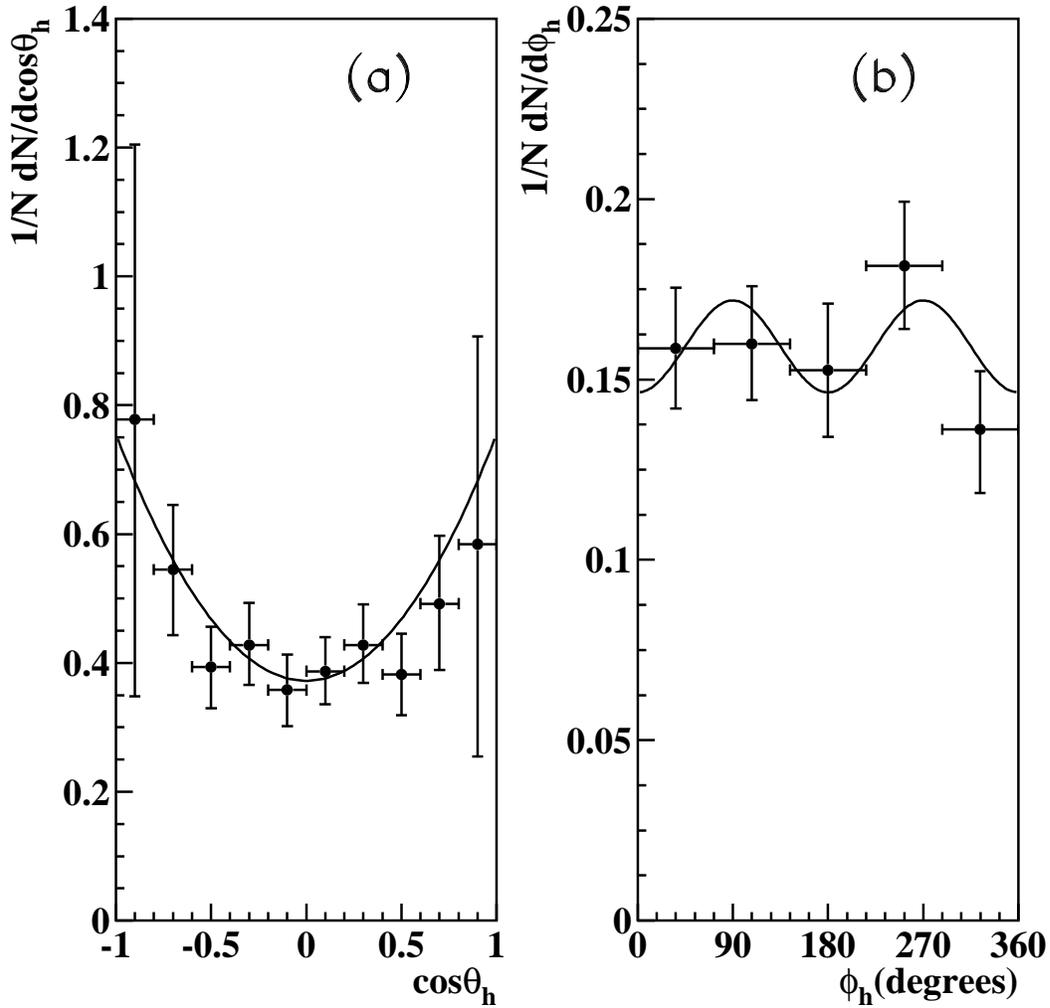,%
       width=15cm%
        }
    \caption{Acceptance corrected decay angular distributions for the \jpsi~
             in the reaction $e p \rightarrow  e J/\psi p$ in the kinematic
             range $40 < W < 140$~GeV. 
             No subtraction of the proton dissociative contribution to
             the sample has been made for the data presented in this
             figure since the angular dependence of the proton
             dissociative and elastic \jpsi~production is assumed to
             be the same.
             The curves are the results of the fits described in the text. 
             The error bars represent the statistical,
             decay-channel-specific errors and common systematic
             errors added in quadrature.
             }
\label{Fig:Heli}
\end{center}
\end{figure}

The decay angular distributions are presented in figure \ref{Fig:Heli}. 
No subtraction of the dissociative contribution has been made for the
distributions presented in figure \ref{Fig:Heli} since it is assumed
that the elastic and dissociative processes have the same angular
dependence.
The distribution of the polar angle of the positive lepton is shown in 
figure \ref{Fig:Heli}a. 
A fit has been used to determine $r^{04}_{00}$.
The result $r^{04}_{00} = -0.01 \pm 0.09$ (which gives $R=-0.01\pm0.09$)
is consistent with SCHC.
The distribution of the azimuthal angle of the positive lepton is
shown in figure \ref{Fig:Heli}b.
The distribution is flat and a fit has been made to determine $r^{04}_{1-1}$. 
Again, the result $r^{04}_{1-1}=-0.08\pm0.07$ is consistent with SCHC.

\section{\bf Summary}
\label{Sect:Conclusion}

The cross section for elastic \jpsi~photoproduction has been measured using the
ZEUS detector at HERA.
A significant rise in the cross section with $W$ has been observed for $W$ in 
the range $40<W<140$~GeV.
The rise in the cross section with $W$ may be parameterised by 
$\sigma_{\gamma p \rightarrow J/\psi p} \propto W^{\delta}$ with 
$\delta = 0.92 \pm 0.14 {\rm (stat.)} \pm 0.10 {\rm (syst.)}$. 
The measured value of $\delta$ is inconsistent with the soft pomeron model.
Models based on the vector dominance model plus the exchange of a
pomeron can be made to describe the data if the effective pomeron
intercept, or the effective pomeron coupling is assumed to depend on
the hard scale in the process. 
QCD based models, which describe the process in terms of the exchange of a
gluon ladder evaluated at leading order or beyond
leading order, are consistent with the data.

The differential cross section ${d\sigma}/{d|t|}$ has been measured and
falls exponentially with $|t|$.
The slope of the exponential has been measured to be 
$4.6\pm0.4^{+0.4}_{-0.6}$~GeV$^{-2}$ in the range $|t|<1$~GeV$^2$.
In geometrical models of vector meson production these results may be
interpreted as indicating that the radius of the \jpsi~is smaller than
that of the $\rho$, $\omega$ and $\phi$ as measured in photoproduction.

The decay angular distributions are consistent with $s$-channel
helicity conservation.

\newpage
\noindent {\Large\bf Acknowledgements}
\vskip 0.5cm
 
We thank the DESY Directorate for their strong support and encouragement.
The experiment was made possible by the inventiveness and the diligent
efforts of the DESY machine group.
The design, construction and installation of the ZEUS detector have
been made possible by the ingenuity and dedicated efforts of many
people from inside DESY and from the home institutes who are not
listed as authors.
Their contributions are acknowledged with great appreciation.

\newpage
%
%--------- REFERENCES -------------
%

%
%--------  Tables:
%
\newpage
\begin{table}[P]
\begin{sideways}\begin{minipage}[b]{\textheight}
\begin{center}
\begin{tabular}{|c|c|c|c|c|c|c|c|} \hline
%          &      &           &       &                                      & 
%       &                                               &                                              \\ 
$W$ Range & Mode & $N_{Sig}$ & \Acce & $\sigma_{ep \rightarrow e J/\psi p}$ (nb) & 
$\Phi_T$ & $\sigma_{\gamma p \rightarrow J/\psi p}$ (nb) & $\sigma_{\gamma p \rightarrow J/\psi p}$ (nb)\\ 
 (GeV)    &      &           &   &            &
                &  & combined \\ \hline
%          &      &           &       &                                      & 
%       &                                               &                                              \\ \hline 
       &                 &          &      &               &        &                            &                            \\
 40-60 &    $e^+e^-$     & $84\pm10$& $0.28$ & $1.23\pm0.14^{+0.17~+0.12}_{-0.21~-~0}$ & 
                                                 0.0411 & $29.9\pm 3.4^{+~4.1~+~2.9}_{-~5.1~-~0}$&$30.4\pm3.4^{+2.9~+~3.2}_{-4.4~-~0}$\\ 
$\langle W \rangle=49.8\pm0.8$
       &                 &          &      &               &        &                            &                            \\
\cline{2-7}
       &                 &          &      &               &        &                            &                            \\
       &$\mu^+\mu^-$ & $48\pm~7$ & $0.23$ & $1.28\pm0.19^{+0.15~+0.13}_{-0.22~-~0}$ & 
                                                   0.0411 & $31.1\pm 4.6^{+3.6~+~3.2}_{-5.4~-~0}$&                            \\
       &                 &          &      &               &        &                            &                            \\ \hline
       &                 &          &      &               &        &                            &                            \\
60-80 & $e^+e^-$         & $98\pm11$& $0.33$ & $1.24\pm0.13^{+0.16~+0.12}_{-0.20~-~0}$ & 
                                                 0.0266 & $46.6\pm 4.9^{+~6.0~+~4.5}_{-~7.5~-~0}$&$42.9\pm4.5^{+4.1~+~4.1}_{-5.6~-~0}$\\
$\langle W \rangle=71.2\pm0.7$
       &                 &          &      &               &        &                            &                            \\
\cline{2-7}
       &                 &          &      &               &        &                            &                            \\
       &$\mu^+\mu^-$     & $61\pm~8$& $0.35$ & $1.05\pm0.14^{+0.13~+0.11}_{-0.16~-~0}$
                                                 & 0.0266 & $39.5\pm 5.3^{+4.9~+~4.1}_{-6.0~-~0}$&                            \\
       &                 &          &      &               &        &                            &                            \\ \hline
       &                 &          &      &               &        &                            &                            \\
80-100 &    $e^+e^-$     & $92\pm10$ & $0.32$& $1.19\pm0.13^{+0.15~+0.12}_{-0.18~-~0}$ & 
                                                 0.0189 & $63.0\pm 6.9^{+~7.9~+6.3}_{-9.5~-~0}$&$57.7\pm5.8^{+5.3~+5.8}_{-6.9~-~0}$\\
$\langle W \rangle=89.6\pm0.7$
       &                 &          &      &               &        &                            &                            \\
\cline{2-7}
       &                 &          &      &               &        &                            &                            \\
       &$\mu^+\mu^-$     &$70\pm~9$ & $0.42$ & $1.01\pm0.12^{+0.14~+0.10}_{-0.14~-~0}$ &
                                                   0.0189 & $53.4\pm 6.3^{+7.4~+~5.3}_{-7.4~-~0}$&                            \\
       &                 &          &      &               &        &                            &                            \\ \hline
       &                 &          &      &               &        &                            &                            \\
100-140    &$e^+e^-$     & $81\pm~9$ & $0.21$& $1.59\pm0.18^{+0.24~+0.16}_{-0.27~-~0}$ & 
                                                 0.0251 & $63.3\pm7.2^{+9.6~+~6.4}_{-10.8~-~0}$&$66.5\pm6.8^{+6.4~+6.8}_{-9.6~-~0}$\\
$\langle W \rangle=121\pm1$
       &                 &          &      &               &        &                            &                            \\
\cline{2-7}
       &                 &          &      &               &        &                            &                            \\
       &$\mu^+\mu^-$     & $87\pm10$& $0.30$ & $1.74\pm0.20^{+0.23~+0.17}_{-0.28~-~0}$ &
                                                   0.0251 & $69.3\pm 8.0^{+9.2~+6.8}_{-11.2~-~0}$&                            \\
       &                 &          &      &               &        &                            &                            \\ \hline
\end{tabular}
    \caption{The results for the integrated \jpsi~photoproduction cross section as a function of $W$.
             $N_{Sig}$ is the number of events after subtraction of the Bethe-Heitler contribution and 
             \Acce~is the acceptance.
             The photon flux $\Phi_T$ is calculated as described in the text and used to calculate the $\gamma p$ cross section,
             $\sigma_{\gamma p \rightarrow J/\psi p}$, from the $ep$ cross section, $\sigma_{ep \rightarrow e J/\psi p}$.
             Cross sections for the individual channels are quoted with the first error being statistical and the second 
             systematic.
             The third error is the error attributed to the model of proton dissociation used for background subtraction and is described
             in the text.
             The combined electron and muon results have been obtained by averaging as described in the text.
             Here the first error contains the combined statistical
             and decay channel specific errors while the second contains all sources of common systematic error.
             The error attributed to the model of proton dissociation is the third error.
             }
\label{Tab:GammaP}
\end{center}
\end{minipage}\end{sideways}
\end{table}
\newpage
\begin{table}[P]
\begin{center}
\begin{tabular}{|l|c|c|c|c|c|c|c|c|}                  \hline
\multicolumn{9}{|c|}{\bf Breakdown of Contributions to the Systematic Error} \\
\multicolumn{9}{|c|}{\it Values are quoted in percent}   \\ 
\multicolumn{9}{|c|}{\it Decay Channel Specific Systematic Errors}                                             \\ \hline 
            & \multicolumn{4}{c|}{Electron Channel}         & \multicolumn{4}{c|}{Muon Channel}       \\ 
            \hline
$W$ bin (GeV)   & {\small 40-60 } & {\small 60-80 } & {\small 80-100 } & {\small 100-140 }
            & {\small 40-60 } & {\small 60-80 } & {\small 80-100 } & {\small 100-140 }   \\ \hline   
{\small Trigger}
            & {\small $^{+7}_{-7}$} & {\small $^{+7}_{-7}$} & {\small $^{+7}_{-7}$} & {\small $^{+7}_{-7}$} 
            & {\small $^{+5}_{-5}$} & {\small $^{+5}_{-5}$} & {\small $^{+5}_{-5}$} & {\small $^{+5}_{-5}$}                 
            \\ \hline 
% & & & & & & & & \\
{\small Event selection}
            & {\small $^{+5.7}_{-4.6}$} & {\small $^{+4.6}_{-4.9}$} & {\small $^{+3.0}_{-4.0}$} & {\small $^{+8.8}_{-3.8}$}         
            & {\small $^{+2.0}_{-6.3}$} & {\small $^{+5.6}_{-4.1}$} & {\small $^{+7.1}_{-0.0}$} & {\small $^{+7.0}_{-2.0}$}    
            \\ \hline
% & & & & & & & & \\                
{\small Pion misidentification}  
            & {\small $^{+~0}_{-1.5}$} & {\small $^{+~0}_{-1.5}$} & {\small $^{+~0}_{-1.5}$} & {\small $^{+~0}_{-1.5}$} &  &   &    &    
            \\ \hline
% & & & & & & & & \\    
{\small Muon chamber} & & & & & & & & \\    
{\small efficiency}  
            &        &        &          &          & {\small $^{+2}_{-2}$} & {\small $^{+2}_{-2}$} 
                                                    & {\small $^{+2}_{-2}$} & {\small $^{+2}_{-2}$}  
            \\ \hline
% & & & & & & & & \\             
{\small Branching ratio}
            & {\small $^{+3.2}_{-3.2}$} & {\small $^{+3.2}_{-3.2}$} & {\small $^{+3.2}_{-3.2}$} & {\small $^{+3.2}_{-3.2}$}  
            & {\small $^{+3.2}_{-3.2}$} & {\small $^{+3.2}_{-3.2}$} & {\small $^{+3.2}_{-3.2}$} & {\small $^{+3.2}_{-3.2}$}
            \\ \hline
% & & & & & & & & \\    
{\bf Subtotal}  
            & {\small $^{+9.6}_{-9.1}$} & {\small $^{+9.0}_{-9.2}$} & {\small $^{+8.3}_{-8.8}$} & {\small $^{+11.7}_{-8.7}$}       
            & {\small $^{+6.6}_{-8.9}$} & {\small $^{+8.4}_{-7.5}$} & {\small $^{+9.5}_{-6.3}$} & {\small $^{+9.4}_{-6.6}$} 
            \\ \hline  
\multicolumn{9}{|c|}{ }                                                                               \\ 
\multicolumn{9}{|c|}{\it Common Systematic Errors}                                               \\ \hline  
$W$ bin (GeV)   & \multicolumn{2}{|c|}{\small 40-60 } & \multicolumn{2}{|c|}{\small 60-80 } & \multicolumn{2}{|c|}{\small 80-100 } 
            & \multicolumn{2}{|c|}{\small 100-140 } \\ \hline
\multicolumn{1}{|l|}{\small Acceptance} & \multicolumn{8}{|c|}{$^{+3}_{-3}$}      \\ \hline 
\multicolumn{1}{|l|}{\small Elastic definition} & \multicolumn{8}{|c|}{$^{+1}_{-3}$}      \\ \hline 
\multicolumn{1}{|l|}{\small Radiative corrections} & \multicolumn{8}{|c|}{$^{+4}_{-4}$}      \\ \hline 
\multicolumn{1}{|l|}{\small Helicity distribution} & \multicolumn{2}{|c|}{$^{+~0}_{-10}$} & \multicolumn{2}{|c|}{$^{+0}_{-8}$}
            & \multicolumn{2}{|c|}{$^{+0}_{-6}$} & \multicolumn{2}{|c|}{$^{+~0}_{-10}$} \\ \hline
\multicolumn{1}{|l|}{\small Proton dissociation} & \multicolumn{8}{|c|}{$^{+6}_{-7}$}      \\ \hline
\multicolumn{1}{|l|}{\small Model of dissociation} & \multicolumn{8}{|c|}{$^{+10}_{~-0}$} \\ \hline
\multicolumn{1}{|l|}{\small $\psi^{'}$ contamination} & \multicolumn{8}{|c|}{ $^{+1}_{-1}$}     \\  \hline
\multicolumn{1}{|l|}{\small Luminosity}  & \multicolumn{8}{|c|}{$^{+1.5}_{-1.5}$} \\ \hline
\multicolumn{9}{|c|}{ }                                                                               \\ 
\multicolumn{9}{|c|}{\it Total Systematic Errors}                                               \\ \hline    
            & \multicolumn{4}{c|}{Electron Channel}         & \multicolumn{4}{c|}{Muon Channel}       \\ 
            \hline
$W$ bin (GeV)   & {\small 40-60 } & {\small 60-80 } & {\small 80-100 } & {\small 100-140 }
            & {\small 40-60 } & {\small 60-80 } & {\small 80-100 } & {\small 100-140 }   \\ \hline   
{\bf Total} & {\small $^{+16.0}_{-16.4}$} & {\small $^{+15.7}_{-15.3}$} & {\small $^{+15.3}_{-14.1}$} & {\small $^{+17.4}_{-16.2}$}       
            & {\small $^{+14.5}_{-16.3}$} & {\small $^{+15.4}_{-14.4}$} & {\small $^{+16.0}_{-12.7}$} & {\small $^{+15.9}_{-15.2}$} 
            \\ \hline  
\end{tabular}
    \caption{The contributions to the systematic errors on the $J/\psi$ photoproduction
             cross section. The contributions to the systematic error are
             divided into {\it Decay Channel Specific Systematic Errors} and 
             {\it Common Systematic Errors} as described in section 8.
             }
\label{Tab:SysErr}
\end{center}
\end{table}
%
%--------  Figures:
%
\newpage
%..  Introduction:
\newpage
\newpage
\newpage
\newpage
%
%--------  The end
%
\end{document}